\PassOptionsToPackage{fleqn}{amsmath}
\documentclass[twocolumn,showpacs,preprintnumbers,amsmath,amssymb]{revtex4}
\usepackage{graphicx}% Include figure files
\usepackage{dcolumn}% Align table columns on decimal point
\usepackage{bm}% bold math
\usepackage{amssymb}
\usepackage{amsmath}

\begin{document}

\title{Growth facets of SrIrO$_3$ Thin Films and Single Crystals}

\author{L. Fruchter$^{1}$,V. Brouet$^{1}$, F. Brisset$^{2}$, H. Moutaabbid$^{3}$,Y. Klein$^{3}$}%
\affiliation{$^a$Laboratoire de Physique des Solides, C.N.R.S., Universit\'{e} Paris-Sud, 91405 Orsay, France}
\affiliation{$^b$SP2M – ICMMO (UMR CNRS 8182), Univ. Paris-Sud, Univ. Paris-Saclay, F-91405 Orsay, France}
\affiliation{$^c$Institut de Min\'{e}ralogie, de Physique des Mat\'{e}riaux et de Cosmochimie (IMPMC), Sorbonne Universit\'{e}, CNRS, IRD, MNHN, 4 place Jussieu 75005 Paris, France}

\begin{abstract}{}
The crystallographic orientation of SrIrO$_3$ surfaces is decisive for the occurrence of topological surface states. We show from density functional theory computations that (001) and (110) free surfaces have comparable energies, and, correspondingly, we experimentally observe that single micro-crystals exhibit both facet orientations. These surfaces are found to relax over typically the length of one oxygen octahedron, defining a structural critical thickness for thin films. A reconstruction of the electronic density associated to tilts of the oxygen octahedra is observed. At the opposite, thin films have invariably been reported to grow along the (110) direction. We show that the interfacial energy associated to the oxygen octahedra distortion for epitaxy is likely at the origin of this specific feature, and propose leads to induce (001) SrIrO$_3$ growth.
\end{abstract}

\maketitle

%%%MAIN TEXT%%%%
\section{Introduction}
Amongst the Ruddlesden-Popper series for iridates, R$_{n+1}$Ir$_n$O$_{3n+1}$ where R= Sr, Ba and n = 1,2,$\infty$, SrIrO$_3$ (n = $\infty$, SIO) occupies a peculiar place. While the n=1,2 compounds are antiferromagnetic insulators with a layered structure, SIO is a three-dimensional, nearly compensated paramagnetic semi-metal\cite{Manca18}. It was proposed that the semi-metallic state is stabilized as bands cross at a topologically-protected line of nodes below the Fermi level\cite{Carter12,Zeb12}. In SIO the Dirac line of nodes is enforced by symmetry in the nonsymmorphic space group \textit{Pbnm}\cite{Armitage18}, and the key-role of \textit{n}-glide operation was stressed as spin-orbit coupling is present\cite{Armitage18,Liu16b}. As a consequence, the crystallographic nature of the surface of practical samples is decisive, when the existence of topological surface states is questioned\cite{Chen15,Chen16,Liu16b}.

Elaboration of bulk samples in the polycrystalline form, using high-pressure synthesis, have been reported up to now only\cite{Longo71,Zhao08}, and epitaxy was used to stabilize the single crystal material. The perovskite structure for SIO makes it compatible with many other perovskite substrates commonly used in film growth. Indeed, SIO thin films have been grown using molecular beam epitaxy\cite{Nie15,Liu16c}, pulsed laser deposition\cite{Moon08,Jang10,Wu13,Liu13,Biswas14,Liu16b,Gruenewald14,Biswas15}, metalorganic chemical vapor deposition\cite{Kim06}, and RF sputtering\cite{Fruchter16}.

While rough structural studies of these films generally rely on the pseudo-cubic structure of the primitive cell of the material, there has been some detailed structural investigations of SIO and the parent ruthenate compound SrRuO$_3$ thin films. It has been reported that SrRuO$_3$  epitaxial layers are grown on (001) SrTiO$_3$ with out-of-plane [110] direction and [001] and [1-10] aligned in-plane\cite{Gan99,Vailionis07,Vailionis08,Vailionis11}. A similar growth orientation was reported for SIO films on (001) SrTiO$_3$, elaborated with PLD, as evidenced by electron diffraction observations\cite{Zhang13}, and for SIO grown on GdScO$_3$, from a complete structure determination\cite{Liu16b}. It was found that the substrate vicinality systematically orients the in-plane orientation of the films, with the [001] direction lying nearest to the vicinal steps one.

In order to gain some understanding on the mechanisms that drive the surface growth selectivity, we have investigated SIO free-standing single crystal surfaces, and computed surface energies and configurations, using \textit{ab initio} density functional theory (DFT).

\section{Experimental}
\subsection{Single crystals}

\begin{figure}
\centering
\includegraphics[height=3.5cm]{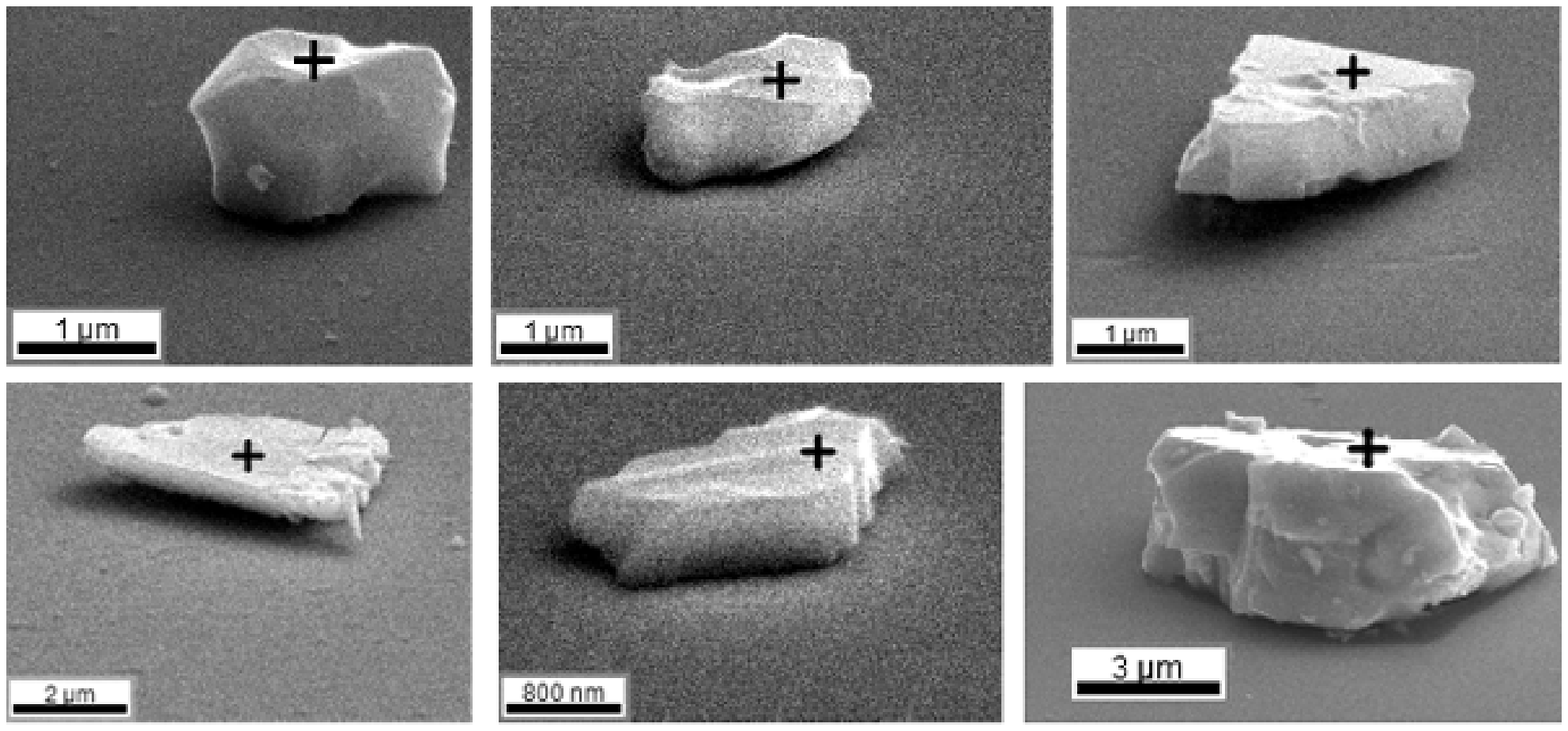}
\caption{A selection of crystals with (110) top surface. The cross indicates the diffraction point.}\label{110_crystals}
\end{figure}

\begin{figure}
\centering
\includegraphics[height=3.5cm]{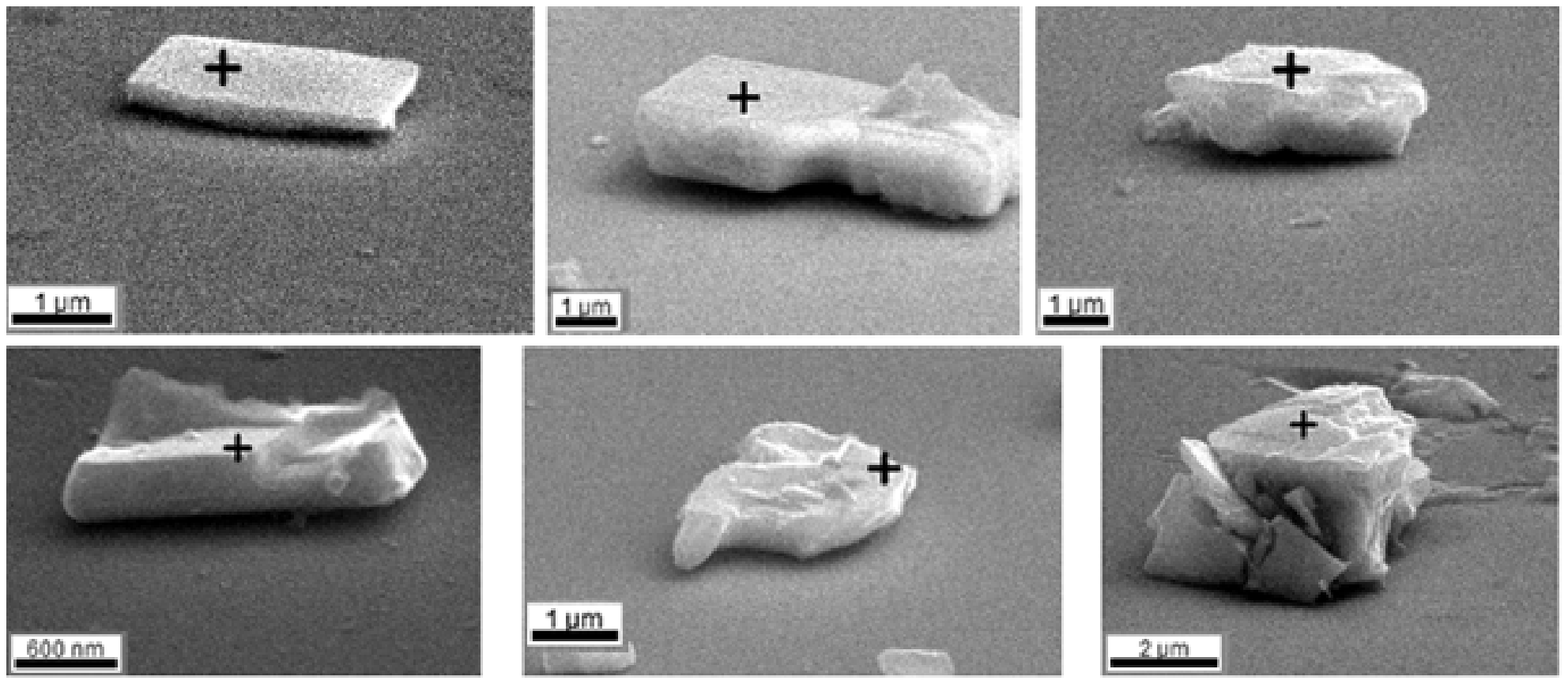}
\caption{Same as in Fig.~\ref{110_crystals}, with (001) top surface.}\label{001_crystals}
\end{figure}

We obtained single crystals from a SIO sintered pellet, grown under high pressure. The high pressure synthesis was realized using a Paris-Edinburgh VX3 hydraulic press, as described in details elsewhere\cite{Klotz04}. In brief, powder of pristine hexagonal SIO was pressed into a 2.9-mm-diameter and 5.5-mm-high pellet and wrapped into a thin platinum foil. The latter was then introduced in a MgO capsule sleeve which was in turn inserted into a graphite cylinder acting as a heater. We applied a quasi-hydrostatic pressure of 5 GPa and a temperature of  1000 $^\circ$C +/- 50 $^\circ$C. The capsule was kept at constant pressure and constant temperature for 30 min and then quenched to room temperature in a few seconds. We then slowly released the pressure at a rate of 0.5 GPa/h. To separate single crystals, the pellet was first finely grounded in dilute polymethyl methacrylate (PMMA) resist. A silicon wafer was then spin-coated, with a drop of this PMMA/SIO mixture. To eliminate the PMMA resist from the surface of the SIO grains, the wafer was finally etched using an oxygen plasma. As can be seen in Figs.~\ref{110_crystals} and \ref{001_crystals}, several single crystals with well defined facets were glued on the wafer, with their upper flat surface parallel to it. Such a configuration, where the surface orientation is approximately known in the electronic diffraction geometry, allowed to determine the surface crystallographic orientation, using Kikuchi line maps patterns.

The patterns from the back scattered electron diffraction (EBSD) were obtained with the wafer normal oriented at 20 deg. from the electron beam, in a  Zeiss Supra 55 VP scanning electron microscope fitted with a Hikari EBSD System from TSL-EDAX\textsuperscript{TM} microscope. We assumed that the single crystals had two parallel surfaces aligned with the wafer (Figs.~\ref{110_crystals} and \ref{001_crystals}), to infer the orientation of the incident beam with respect to the crystal surface. The accelerating voltage was 20 kV and the current a few nA. The Kikuchi maps were analyzed using OIM\textsuperscript{TM} software.

\subsection{DFT computations}

In order to compute the surface energies, we have performed DFT computation of the electronic energy of periodic slabs, using the full-potential linearized augmented plane wave (FP-LAPW) method, as implemented in the Wien2k software\cite{Blaha90}.

\begin{figure}
\centering
\includegraphics[height=5cm]{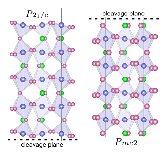}
\caption{Symmetric cells with (001) IrO$_2$/SrO free surface, obtained from the cleavage of $N$ = 4.5 bulk unit cells. The black lines delimit the unit cells.}\label{cells}
\end{figure}

As pointed out by Heifets et al\cite{Heifets01}, the surface energy is the sum of the cleavage and relaxation energies. One may then virtually cleave a bulk cell made of $N$ structure unit cells, and compute the difference in energy between the cleaved slabs and the bulk material, for both the unrelaxed and relaxed structures. As for SrTiO$_3$, we expect that, due to the covalent nature of the Ir-O bonding, as compared to ionic character of the Sr-O one, the SrO-IrO$_2$ stacking is weakly polar with a negative charge in the SrO plane, although formal valences assign the compound to Tasker's classification type I\cite{Tasker}. In order to impede the formation of an electrostatic macroscopic dipole, which could induce a spurious reconstruction of the surface, it is preferable to create two symmetric slabs (\textit{i.e.} with identical terminations at both ends), each of them terminated with different atomic planes\cite{Piskunov05}. These terminating layers may be SrO or IrO$_2$ (Fig.~\ref{cells}).

To investigate the (001) surface, we built couples of slabs consisting of the stacking along the \textit{c}-axis of SrO-IrO$_2$-...-IrO$_2$-SrO and IrO$_2$-SrO-....-SrO-IrO$_2$ (001) layers. Depending on the number of unit cells contained in the original bulk cell ($N$), the slabs could be made of $N/2$ unit cells ($N$ even), or $(N-1)/2$ unit cells ($N$ odd). While the bulk structure has the \textit{Pbnm} symmetry\cite{Zhao08,Puggioni16} (No. 62 - 8 symmetry operations), the cleaved slabs have a lower orthorhombic symmetry, obtained as the maximal non isomorphic subgroup of this group with the preserved symmetry elements: 
\\- \textit{P2$_{1/c}$} (No. 14 - 4 symmetry operations, including inversion center) for $N$ even, IrO$_2$-terminated cells, and $N$ odd, SrO-terminated slabs,
\\- \textit{Pmc2$_{1}$} (No. 26 - 4 symmetry operations, including a \textit{c}-axis mirror plane) for $N$ even, SrO-terminated cells, and $N$ odd, IrO$_2$-terminated slabs.\\In the case of (110) surface, the [110] direction makes an angle $\gamma$ = 89.6 deg. from the (110) plane. As a consequence, the slabs have a lower, monoclinic symmetry, with one mirror symmetry, \textit{Pm} (No. 6). The unit cell is also larger by $\approx$ $\sqrt{2}$ in the basal plane, with unit cell parameter 7.98 \AA along the [-110] direction, and 7.948 \AA along the [001] one.

The vacuum between each cell was chosen 10 \AA wide. We used for the bulk structure the one obtained by relaxing both the cell parameters and the internal coordinates. As can be seen from Table~\ref{tab:table1}, the fully relaxed structure has oxygen octahedra slightly more tilted than found in the literature, and larger cell parameters. To be consistent with our free-surface cell relaxation procedure, we used this relaxed structure to build the cleaved slabs. For the consistency, also, between energy and structural relaxation computations, muffin-tin radii were systematically reduced by 8\% for all computations. Structural relaxations were performed down to residual force 0.2 mRy/a.u. We found that a $k$-mesh with a number of $k$-points as small as 500 was sufficient to insure the energy convergence of the cells, with accuracy well within the measured energy differences.

\begin{table}[h]
\small
\caption{\ SrIrO$_3$ structural parameters}
\label{tab:table1}
\begin{tabular*}{0.48\textwidth}{@{\extracolsep{\fill}}lllll}
\hline
 & Ref.~\cite{Zhao08} & Ref.~\cite{Puggioni16} & Ref.~\cite{Blanchard14} & This work\\
\hline
a (\AA) & 5.5617 & \textnormal{"} & 5.6008 & 5.622\\
b (\AA) & 5.5909 & \textnormal{"} & 5.5712 & 5.663\\
c (\AA) & 7.8821 & \textnormal{"} & 7.8960 & 7.948\\
Ir-O$_{bas.}$-Ir (deg.)$^a$ & 153.5 & 154.0 & 156.2 & 153.8\\
Ir-O$_{ap.}$-Ir (deg.)$^a$ & 156.6 & 154.5 & 156.9 & 153.3\\
E/unit cell (eV)$^b$ & 0.22 & 0.73 & 0.31 & 0\\
\hline
\end{tabular*}
\footnotesize{$^a$ O$_{bas./ap.}$ = basal-plane / apical oxygen,$^b$ referenced to this work}
\end{table}

Given the total electronic energy for cells as defined above, the surface energy may be computed using:

\begin{eqnarray*}\label{surfenergy}
E(\textnormal{\small SrO/IrO$_2$\normalsize}) = E_{cleavage} + E_{rel}(\textnormal{\small SrO/IrO$_2$\normalsize}) \\ 
E_{cleavage} = 1/4 [E_{unrel}(\textnormal{\small SrO\normalsize}) + E_{unrel}(\textnormal{\small IrO$_2$\normalsize}) - N E_{bulk}] \\
E_{rel}(\textnormal{\small SrO/IrO$_2$\normalsize}) = 1/2 [E_{rel}(\textnormal{\small SrO/IrO$_2$\normalsize})-E_{unrel}(\textnormal{\small SrO/IrO$_2$\normalsize})]
\end{eqnarray*}

where $E_{rel/unrel}$(\textnormal{\small SrO/IrO$_2$\normalsize}) is the total electronic energy for the relaxed/unrelaxed cell, with SrO/IrO$_2$ termination, and $E_{bulk}$ is the electronic energy for one (relaxed) bulk unit cell.

\section{Results and discussion}

\begin{figure}
\centering
\includegraphics[height=4cm]{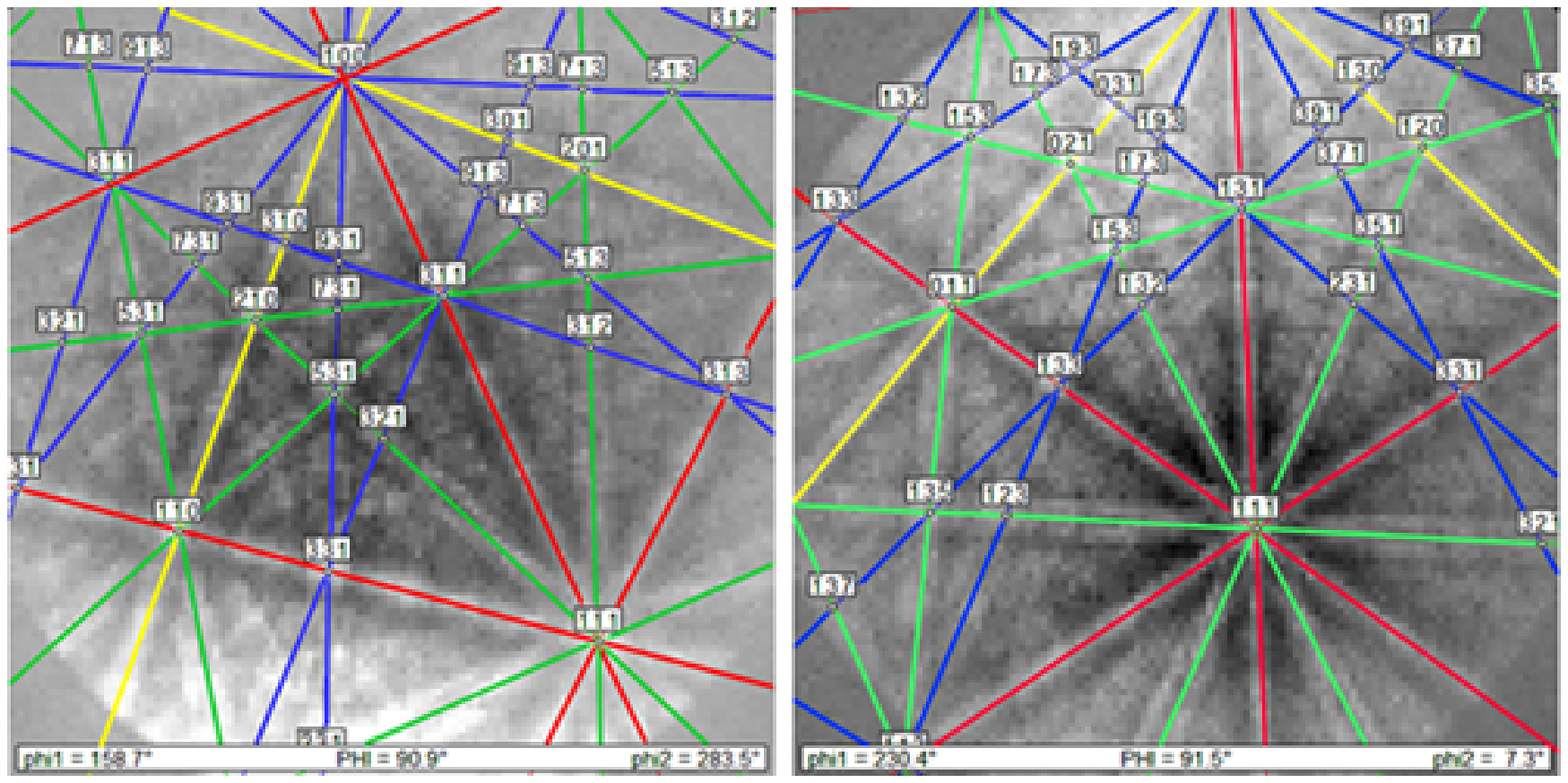}
\caption{Indexed Kikuchi diagrams for bottom right crystals in Fig.~\ref{110_crystals} (left) and \ref{001_crystals} (right). The labels are the crystal directions at the intersection of lattice planes.}\label{Kikuchi}
\end{figure}

\begin{figure}
\centering
\includegraphics[height=4cm]{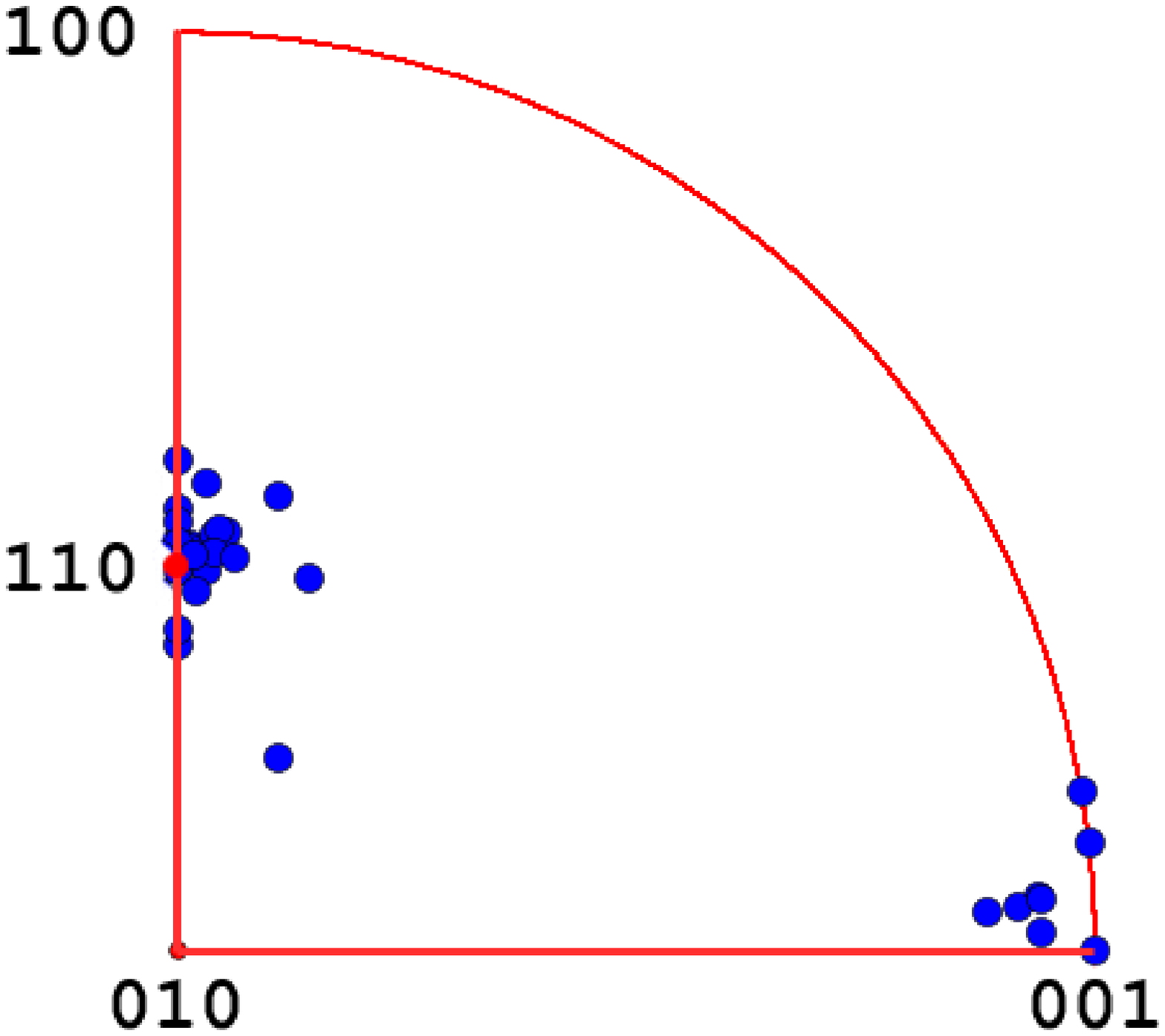}
\caption{Inverse pole figure showing the orientation of the facet parallel to the substrate for 32 crystals (indices for \textit{Pbnm} notation).}\label{distribution}
\end{figure}

Very well contrasted Kikuchi maps were obtained from the selected crystals, indicating clean, metallic surfaces, with a well defined crystallographic orientation (Fig.~\ref{Kikuchi}). The Kikuchi diffraction lines result from the diffraction by the atomic planes of the diffused electron beam (a mechanism similar to diffused X-ray diffraction for the Kossel lines), that is typically from a few tens nanometers thick layer from the surface. The pole figure for a collection of grains (Fig.~\ref{distribution}) only shows two dominant orientations for the top surfaces, with 75\% of them in the (110) orientation, and the rest in the (001) one. Considering the degeneracy of the ($\pm$1$\pm$10) surfaces and (00$\pm$1) ones for the Kikuchi maps, the expectation ratio is 2 for equiprobable occurrence of such orientations. Due to the limited number of available samples (32), the excess of our (110) observations cannot invalidate the hypothesis of an equiprobable occurrence for each orientation with more than $\approx$ 60 \% confidence. So, we consider our observations to be compatible with equiprobable occurrence within a reasonable error bar.

\begin{figure}
\centering
\includegraphics[height=4cm]{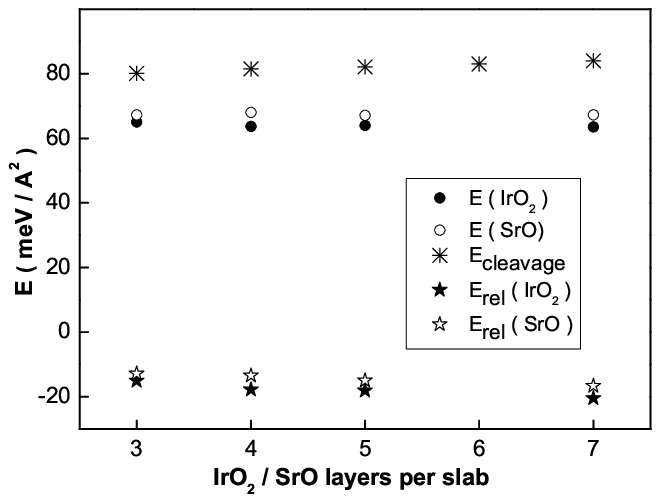}
\caption{Energies for (001) surfaces. The abscissa is the number of IrO$_2$ (alt. SrO) layers in the IrO$_2$-terminated (alt. SrO) slab.}\label{E001}
\end{figure}

To compute the surface energies, we first looked for possible finite size effects. Fig.~\ref{E001} displays the surface energy for (001) slabs of different thicknesses. Energies depend very weakly on the slab thickness, down to slabs with only 3 IrO$_2$/SrO layers: this is an indication that the electronic and structural properties relax over quite a short distance from the surface, as will be seen below. The relaxation energy contributes to typically 30 \% of the total energy, as is commonly the case. The SrO surface energy is systematically larger than the IrO$_2$ one, by about 3-7 \%. Similar, but less extensive computations -- due to the lack of symmetry, which considerably increases the computation time and the convergence accuracy -- for (110) slabs were performed. As may be seen from table~\ref{tab:table2}, surface energies are comparable within error bars to the ones for (001) slabs. This is in agreement with our observation of comparable occurrence for (001) and (110) facets for single crystals. 

\begin{table}
\small
\caption{\ Relaxation and total surface energies for (001) and (110) slabs (meV/\AA$^2$)}
\label{tab:table2}
\begin{tabular*}{0.48\textwidth}{@{\extracolsep{\fill}}lllll}
\hline
 & E$_{rel}$(001)& E(001) & E$_{rel}$(110) & E(110) \\
\hline
IrO$_2$& -17.9 $\pm$ 0.1 & 64.1 $\pm$ 0.7 & -14 $\pm$ 0.4 & 66 $\pm$ 4 \\
SrO & -14.5 $\pm$ 0.1 & 67.5 $\pm$ 0.7& -18 $\pm$ 1 & 63 $\pm$ 4 \\
\hline
\end{tabular*}
\end{table}

\begin{figure}
\centering
\includegraphics[height=4cm]{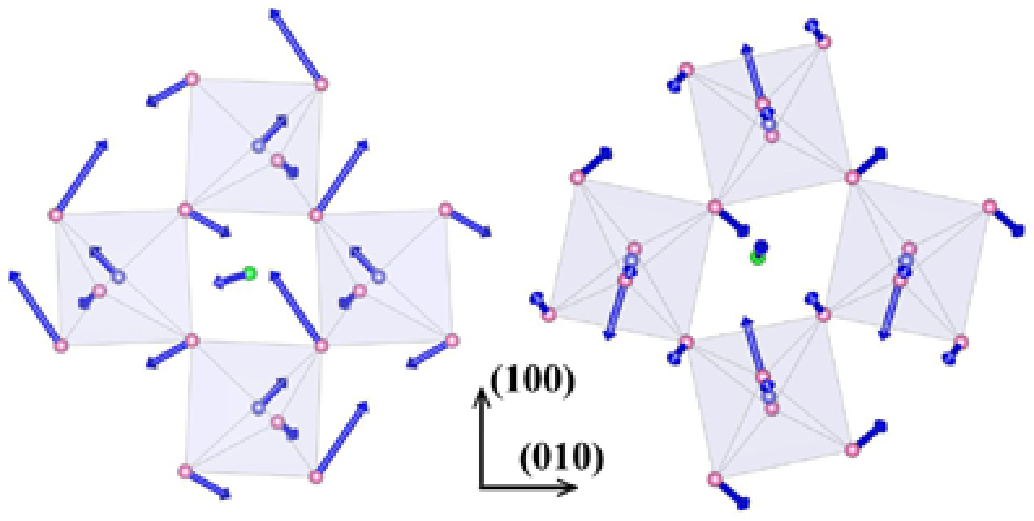}
\caption{Relaxation of the top layers for a (001) IrO$_2$-terminated (left) and SrO-terminated slab (right) (top view). The length of the arrows is proportional to the displacement amplitude.}\label{relax}
\end{figure}

\begin{figure}
\centering
\includegraphics[height=4cm]{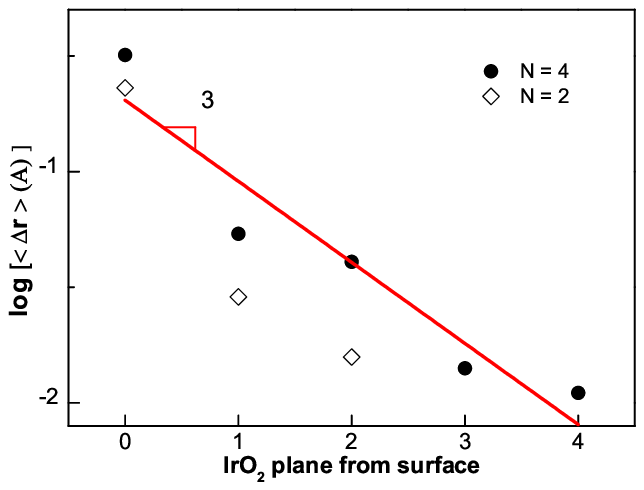}
\caption{Average atom displacement in an IrO$_2$ layer, for a $N$=8 IrO$_2$-terminated slab (the surface layer is indexed 0). The line is an approximation for an exponential decay of the $N$ = 4 data.}\label{decay}
\end{figure}

\begin{figure}
\centering
\includegraphics[height=3cm]{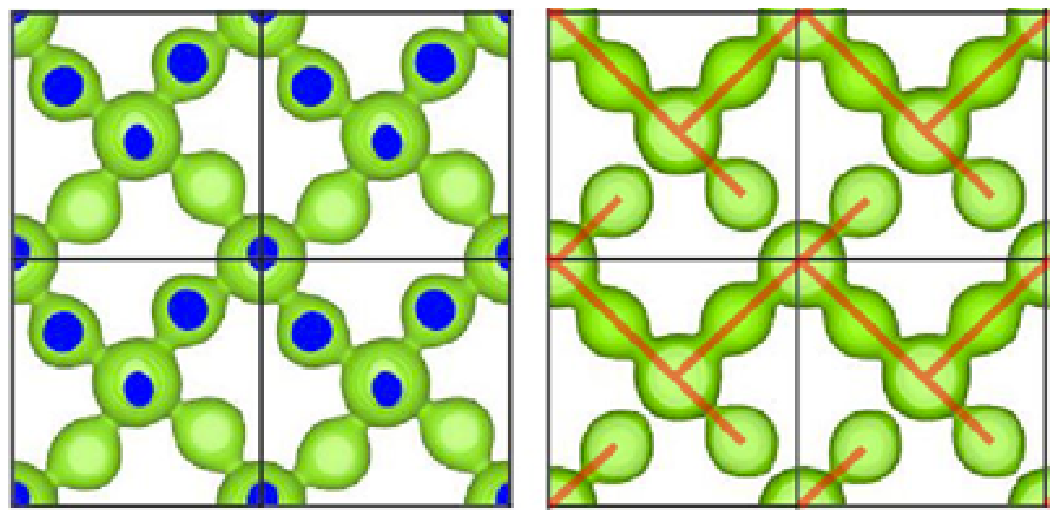}
\caption{Iso-electronic density surface in the IrO$_2$ plane. a: bulk (the blue patches are the cut of the 3D surface). b: surface layer of IrO$_2$-terminated (001) slab (the red lines highlight the paths with the largest electronic density). Black lines are for the unit cell.}\label{clouds}
\end{figure}

The top layers of a relaxed (001) IrO$_2$-terminated slab are displayed in Fig.~\ref{relax}. It may be seen that the oxygen octahedra of the vacuum-interface layer rotate around the \textit{c}-axis, so as to increase the Ir-O$_{bas}$-Ir angle strongly (+8.3 deg.), while they tilt from this axis, so as to decrease the Ir-O$_{ap}$-Ir angle (-3.5 deg.). The structural relaxation vanishes in the bulk, with a typical length which may be defined from the approximate exponential decay of the averaged atomic displacements in each layer. As can be seen in Fig.~\ref{decay}, this length is as short as about 1.3 oxygen octahedra: structural relaxation effects should manifest themselves in the electronic properties of films thinner than this limit, as well as for electronic probes below this depth (such as photo-emission). Indeed, the inspection of the electronic iso-electronic density surfaces for the top IrO$_2$ layer reveals that the tilt of the oxygen octahedra tends to disrupt the electronic cloud at some Ir-O-Ir bonds, where this bound has the largest bending (154.5 deg. for the weak bounds, as compared to 162.1 deg. for the strong ones in Fig.~\ref{clouds}b). This results in the formation of chains along the \textit{b}-axis, which illustrates the lower local symmetry of the surface, whereas uniform Ir-O-Ir angles in the bulk insures equivalent bonding in the IrO$_2$ plane (Fig.~\ref{clouds}a)). This adds a free surface structural limit, in addition to electronic confinement effects and epitaxial constraint ones in thin films\cite{Groenendijk16,Schultz17}. Relaxed (001) SrO-terminated slabs display smaller and opposite displacements (Fig.~\ref{relax}): the octahedra rotation about the \textit{c}-axis increases (-0.7 deg. for the Ir-O$_{bas}$-Ir angle), while the tilt from this axis decreases (+2.4 deg. for the Ir-O$_{ap}$-Ir angle).

\begin{figure}
\centering
\includegraphics[height=4cm]{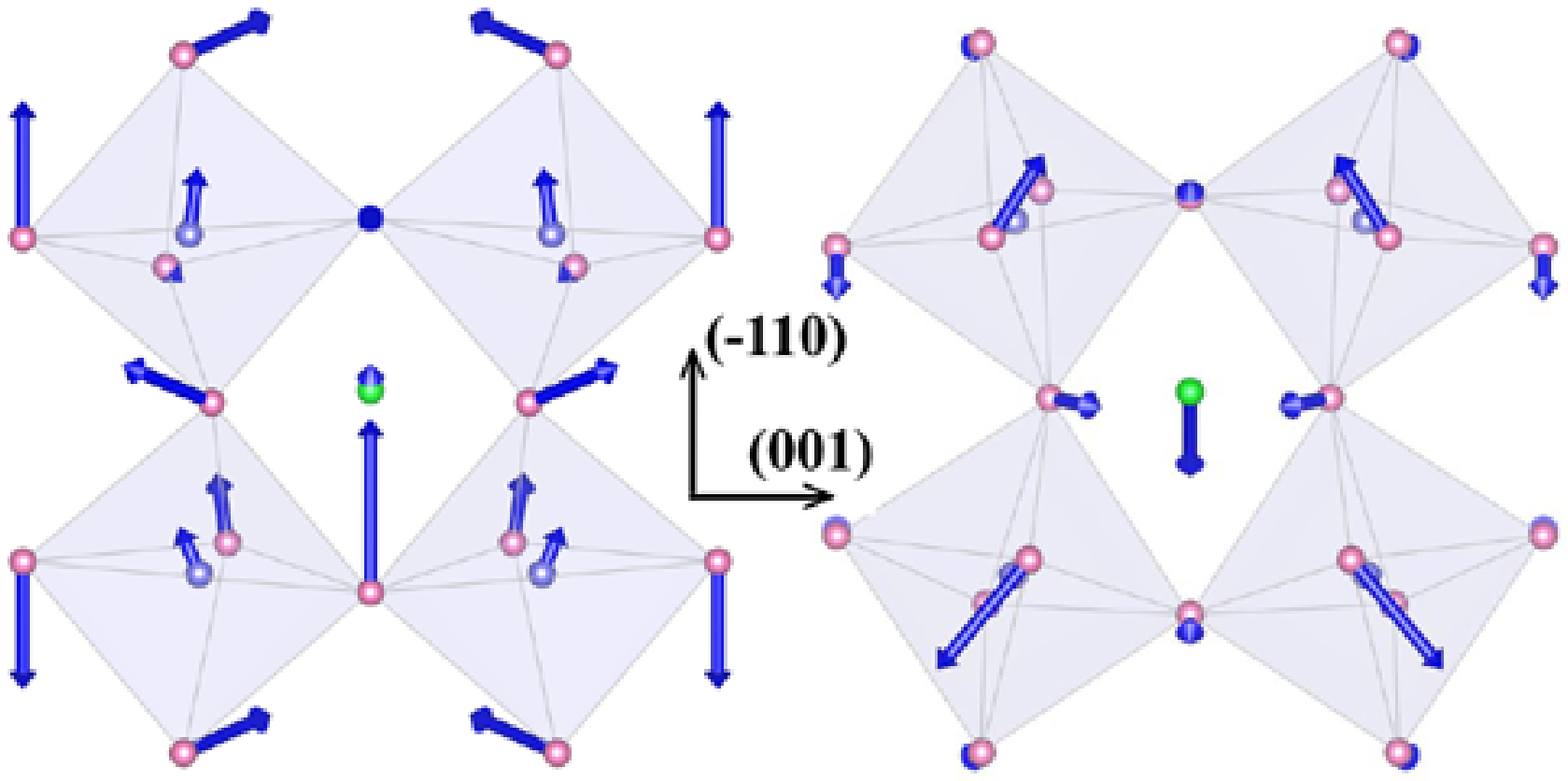}
\caption{Relaxation of the top layers for a (110) IrO$_2$-terminated (left) and SrO-terminated slab (right). (top view)}\label{relax2}
\end{figure}

\begin{figure}
\centering
\includegraphics[height=3cm]{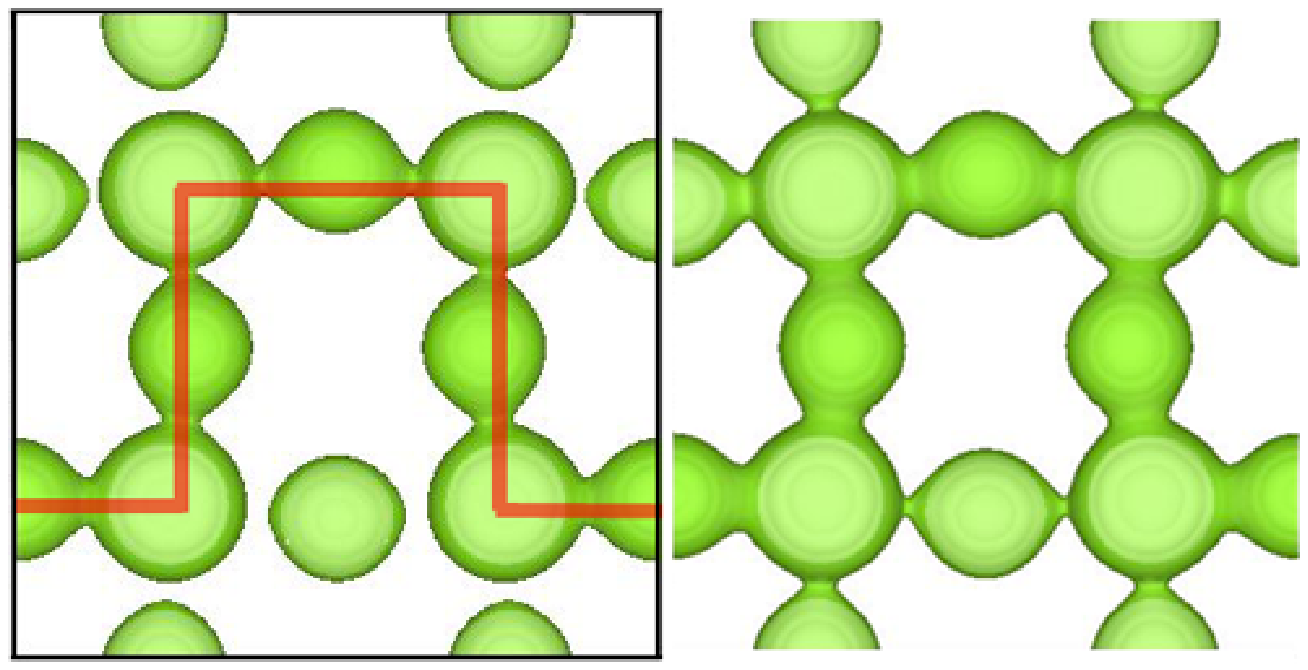}
\caption{Iso-electronic density surface ($d$ = 0.12 (a.u.)$^{-3}$) in the IrO$_2$ plane for a IrO$_2$-terminated (110) slab (the red lines highlight the paths with the largest electronic density). The right panel displays the surface with the same density as in Fig.~\ref{clouds}, $d$ = 0.1 (a.u.)$^{-3}$.}\label{clouds2}
\end{figure}

Relaxed (110) surfaces show similar trends as for (001) surfaces: the IrO$_2$-terminated slabs display an increase of the Ir-O$_{bas}$-Ir angle and a decrease Ir-O$_{ap}$-Ir angle, while SrO-terminated slabs display opposite rotations. An essential difference in this case is due to the presence of the (001) mirror plane, which imposes two basal oxygen to act as pivots with respect to \textit{c}-axis rotations (Fig.~\ref{relax2}). As for the (001) slabs, the rotation of the oxygen octahedra disrupts the electronic cloud, inducing in this case chains along the (001) direction (Fig.~\ref{clouds2}). However, due to the smaller rotations, the electronic constrictions are less marked in this case.

An alternative to quantify the octahedra rotations is to generalize Glazer's rotation angles\cite{Glazer72} to the slab surface (assuming that the surface layer is the one of a bulk crystal with all symmetry elements, and neglecting octahedra distortions). The rotation angles are shown in Table~\ref{glazer}. These observations may be qualitatively understood considering the extra charge brought by the surface termination. As the charge on the Iridium site is increased (IrO$_2$-terminated surface) or decreased (SrO-terminated surface), the Ir-O bond length is expected to increase (alt. to decrease). A direct consequence for this is a smaller (alt. larger) tilt of the octahedra along the axis perpendicular to the IrO$_2$ plane.

\begin{table}
\small
\caption{\ Glazer's rotation angles for the relaxed slabs surface layer (deg.)}
\label{glazer}
\begin{tabular*}{0.48\textwidth}{@{\extracolsep{\fill}}llll}
\hline
 & $\theta_{[100]}$ & $\theta_{[010]}$ & $\theta_{[001]}$ \\
\hline
bulk & 9.8 & 9.8 & 8.6\\
(001)-IrO$_2$ & 11.5 & 11.5 & 16.8 \\
(001)-SrO & 9.4 & 9.4 & 16.8 \\
(110)-IrO$_2$ & 13.4 & 6.7 & 17.2 \\
(110)-SrO & 9.7 & 8.0 & 10.7 \\
\hline
\end{tabular*}
\end{table}

It must be stressed that the above computations are made within the constraint of the given slab cell periodicity and maximal symmetry. While the system does relax towards a local energy minimum in the phase space (which cannot be entirely explored with Wien2k package, which relies on a steepest descent algorithm), we implicitly discarded superstructures and lower symmetries. More stable configurations, with a different periodicity and/or a lower symmetry, may exist. However, the relaxation of a (001) slab with a doubled unit cell (2\textit{a},2\textit{b}), and no crystal symmetry yielded the same configuration as in Fig.~\ref{relax}. So, the relaxed surface appears robust against periodicity and symmetry conditions. This is an indication, also, that symmetry elements needed for topological surface states can be preserved.

Finally, we would like to discuss the consequences of the above computations for the growth of SIO thin films. One generally expects that the free surface energy, and the elastic one associated to the epitaxy, can be decisive for the growth orientation of thin films. As we have seen, the former does not discriminate in favor of one or the other orientation. Considering the elastic energy associated to the material strain, we notice that the (001) epitaxy distinguishes from the (110) one, as it requires only normal strains along the orthorhombic crystal axis, while (110) epitaxy requires both a normal strain along the [001] direction, and a shear strain in the (110) plane.
We have evaluated the stiffness coefficients for SIO, by computing the energy of strained structures\cite{Mehl93}. We find: $C_{11}$ $\approx$ $C_{22}$ $\approx$ $C_{33}$ $\approx$ 260 GPa, $C_{13}$ $\approx$ $C_{23}$ $\approx$ $C_{12}$ $\approx$ 120 GPa, and for the shear coefficient: $C_{66}$ $\approx$ 90 GPa. This evidences the nearly cubic character of the structure. In the framework of the generalized Hooke's law for the linear theory for small deformations, and for a cubic material, it may be shown that the elastic energy, $W$, for the (001) epitaxy is, assuming that the film relaxes freely along the [001] direction:

\begin{equation}
W / \varepsilon^2 = C + C' -2 \,C'^2/C
\label{001energy}
\end{equation}

where $\varepsilon$ is the strain in the substrate plane, assumed identical in the [100] and [010] directions, $C$ = $C_{11}$ = $C_{22}$ = $C_{33}$, and $C'$ = $C_{12}$ = $C_{13}$ = $C_{23}$. For the (110) epitaxy, assuming also identical strain in the [001] and [1$\bar{1}$0] directions, we have:

\begin{equation}
W / \varepsilon^2 = C\, (1+2\, A^2)/2 - C'\,A + 2\, C_{66} \, (1+A)
\label{110energy}
\end{equation}

where $A$ = $C'/(C+C')$. Using the evaluations given above, the stiffness for the (001) growth, $W / \varepsilon^2$, is found 270 GPa from Eq.~\ref{001energy}, to be compared with the one for the (110) growth, 360 GPa, from Eq.~\ref{110energy}. Thus, we find that it is not energetically favorable for stressed films to involve a shear strain, as is the case for the (110) growth, rather than only normal strains, as for the (001) growth. However, the difference in elastic energy between the two orientations is quite small, as compared to typical surface energies. Using the above estimates, a 1 \% mismatch for a one-unit cell thick film yields $\Delta W \approx$ 0.5 meV/\AA$^2$, so we do not expect this contribution to be pertinent, for usual mismatch values.

So, neither a difference in the free surface energy, nor one in the elastic energy associated to the epitaxial constrain, can account for the observed orientation of epitaxial thin films. We have not yet considered, however, the contribution of the epitaxial interface to the film energy, which may be decisive for the orientation of the first layer, and of the subsequent ones. Beyond matching of the film and substrate pseudo-cubic cells, which is usually crudely considered for epitaxy (and obviously not selective for the film orientation), the tilting of the oxygen octahedra required at the interface is potentialy a large, orientation-dependent, contribution to the interface energy. The evaluation of this contribution requires to investigate bilayers, which is beyond the scope of this paper. However, we may roughly compare substrates, using the displacement of the apical oxygen atom, needed to match a given substrate, once the material is strained to accommodate the pseudo-cubic cells. As may be seen in Table~\ref{tab:table3}, amongst the best matched substrates for SIO growth, SrTiO$_3$ shows the lowest selectivity with this criterion, while there is a large difference in favor of the (110) growth on scandates substrates with the usual (110) orientation. We included LaMnO$_3$, although it is not a common substrate. Indeed, (001) LaMnO$_3$ is close to the pseudocubic form, with Mn-Mn angle 90.02 deg. (while Ir-Ir = 90.3 deg. for (001) SIO), allowing for an alternative to the selective growth of (001) SIO, as seen in Table~\ref{tab:table3}. (001) DyScO$_3$ and (001) GdScO$_3$, which show a favorable selectivity, are quite far from the pseudocubic form (respectively, Sc-Sc = 92.8 and 92.7 deg.), but a large descrepancy between the pseudocubic cells may be acceptable, as is the case for TbMnO$_3$ (Mn-Mn = 95.7 deg.) grown on DyScO$_3$\cite{Daumont09}.\newline

\begin{table}
\caption{\ Root mean square displacement of apical oxygen for SrIrO$_3$ epitaxially grown on different substrates (\AA). Favorable situations for (001) growth are highlighted}
\label{tab:table3}
\begin{tabular*}{0.48\textwidth}{@{\extracolsep{\fill}}lll}
\hline
 substrate & (001) SrIrO$_3$& (110) SrIrO$_3$ \\
\hline
(100) SrTiO$_3$ & 0.41 & 0.46 \\
(110) GdScO$_3$ & 0.57 & 0.22 \\
(110) DyScO$_3$ & 0.94 & 0.22 \\
\textbf{(001) GdScO$_3$} & \textbf{0.37} & \textbf{0.65} \\
\textbf{(001) DyScO$_3$} & \textbf{0.39} & \textbf{0.65} \\
(110) LaMnO$_3$ & 0.38 & 0.08 \\
\textbf{(001) LaMnO$_3$} & \textbf{0.04} & \textbf{0.75} \\
\hline
\end{tabular*}
\end{table}

\section{Conclusions}
In summary, we have shown that (001) and (110) SrIrO$_3$ surfaces have comparable surface energies, which is confirmed by the observation that single micro-crystals exhibit both facet orientations with comparable probability. The atomic displacements at the surface were found to relax in the bulk over typically one oxygen octahedron, and to determine electronic channel patterns that run on the cleaved surfaces. For films, we found that usual strains cannot induce a preferential surface orientation, and we suggest that the epitaxy is solely responsible for the observed (110) orientation.

\section*{Conflicts of interest}
There are no conflicts to declare.

\section*{Acknowledgements}
We thank Doroth\'ee Colson and Anne Forget from SPEC-CEA-CNRS for providing us with the monoclinic SrIrO$_3$ material.
We acknowledge support from the Agence Nationale de la Recherche grant "`SOCRATE"' (ANR-15-CE30-0009-01), and LabEx PALM ReflectX (ANR-10-LABX-0039-PALM).

% Non-BibTeX users please use

\end{document}